# Nonlinear Dynamics, Magnitude-Period Formula and Forecasts on Earthquake

*Yi-Fang Chang*
*Department of Physics, Yunnan University, Kunming 650091, China*
(E‐mail: yifangchang1030@hotmail.com)

**Abstract**

Based on the geodynamics, an earthquake does not take place until the momentum-energy excess a faulting threshold value of rock due to the movement of the fluid layer under the rock layer and the transport and accumulation of the momentum. From the nonlinear equations of fluid mechanics, a simplified nonlinear solution of momentum corresponding the accumulation of the energy could be derived. Otherwise, a chaos equation could be obtained, in which chaos corresponds to the earthquake, which shows complexity on seismology, and impossibility of exact prediction of earthquakes. But, combining the Carlson-Langer model and the Gutenberg-Richter relation, the magnitude-period formula of the earthquake may be derived approximately, and some results can be calculated quantitatively. For example, we forecast a series of earthquakes of 2004, 2009 and 2014, especially in 2019 in California. Combining the Lorenz model, we discuss the earthquake migration to and fro. Moreover, many external causes for earthquake are merely the initial conditions of this nonlinear system.

I. **Introduction**

The earthquakes are very complex nonlinear phenomena, and many theories and some phenomenological formulas about this have been proposed. At present the nonlinear seismology is an exciting direction of the development. Some concepts, a phenomenological description of fractals, and propagation and interaction on the seismic wave in the nonlinear media have been discussed [1-4]. Carlson, Langer, et al. [5-8], presented the Burridge-Knopoff block-and-spring model of an earthquake fault, and discussed basic properties, predictability, etc., of the model.

For the forecasts of earthquakes, Kiremidjian, et al. [9] presented a stochastic slip-predictable model based on Markov renewal theory for earthquake occurrences. Borodich [10] described some renormalization schemes for earthquake prediction, which can be reduced to a power-law or the log-periodic approximation of the regional seismic-activity data. Harris [11] summarized more than 20 scientifically based predictions made before the 1989 Loma Prieta earthquake for a large earthquake that might occur in the Loma Prieta region. The predictions geographically closest to the actual earthquakes primarily specified slip on the San Andreas fault northwest of San Juan Bautista. He discussed forecasts of earthquake in 1989 California. Marzocchi, et al. [12] provided insights that might contribute to better formally defining the earthquake‐forecasting problem, both in setting up and in testing the validity of the forecasting model, and found that the forecasting capability of these algorithms is very likely significantly overestimated. Helmstetter, et al. [13] developed a time-independent forecast for southern California by smoothing the locations of magnitude 2 and larger earthquakes, and using small $m \geq 2$ earthquakes gives a reasonably good prediction of $m \geq 5$ earthquakes.

The geodynamics [14] combines the known mantle convection hypothesis, the magma intrusion theory, the phase change theory, and the faulting mechanism, earthquake should be caused by a horizontal fluid layer in a gravitational field that is heated from within and cooled from above, this mantle with very large viscosity moves slowly, and those accompanying momentum-energy transported and accumulated. When these are in excess of a faulting threshold value of rock, a phase transition arises, and an earthquake occurs with the energy releases.

Based on a general geodynamics, we present a simplified and fundamental nonlinear dynamical theory on the earthquake, and obtain the magnitude-period formula of the earthquake combining the Carlson-Langer model, so some forecasts can be calculated quantitatively. Further, the theory may be a basis for the development.

## II. Nonlinear dynamical system of earthquake

It is often convenient to think of continents as blocks of wood floating on a sea of mantle rock [14]. The Carlson-Langer model [5-8] consists of a uniform chain of blocks and springs pulled slowly across a rough surface, in which the nonlinear friction force depends only on the velocity of the block. In this paper we extend the fundamental equations to general nonlinear ones of fluid mechanics in which the conservation of momentum-energy exists, so they are more general equations of magnetohydrodynamics. The equations of momentum conservation are

$$\rho\frac{DV}{Dt} = \rho(\frac{\partial V}{\partial t} + V\nabla V) = F_0 - gradp + f. \tag{1}$$

The equations of energy conservation are

$$\frac{D}{Dt}(\varepsilon + \frac{1}{2}\rho v^2) = vF_0 - vgradp + vf - pdivV + \Phi + Q. \tag{2}$$

where for Newtonian fluid the frictional force is

$$f = \eta\Delta v + (\zeta + \frac{\eta}{3})graddivV. \tag{3}$$

If the rotation of the Earth is concerned in magnetohydrodynamics, the external force will be

$$F_0 = \rho_e E + \frac{\rho_e}{c}V \times H - 2\rho\Omega \times V + \rho R\Omega^2. \tag{4}$$

In these equations there are a few variables, their solutions are very difficult.

In order to simplify the above case, firstly, the electromagnetic phenomena and the rotation of the Earth are neglected, and assume that the pressure gradient $gradp=C''$ which is independent of the velocity V, so Eqs. (1) turn into

$$\frac{\partial V}{\partial t} + V\nabla V = -\frac{C''}{\rho} + \frac{\eta}{\rho}\Delta V + \frac{3\zeta + \eta}{\rho}graddivV. \tag{5}$$

This is a nonlinear partial differential equation of the velocity V. By using the way similar to the soliton solution, let $\xi = \alpha$ $(x+y+z-ut)$, so Eq. (5) becomes an ordinary differential equation

$$(v - u)v' = -\frac{C''}{\rho\alpha} + \alpha bv''. \tag{6}$$

where $b = (4\eta/3 + \zeta)/\rho$. If $C'' \neq 0$, the equation is

$$\frac{dv}{d\xi} = \frac{1}{2\alpha b}(v^2 - 2uv) + C'. \tag{7}$$

$$v = u + b\sqrt{B}cth(-\frac{1}{2}\sqrt{B}\xi + C), \tag{8}$$

for $(u/\alpha b)^2 - (2C'/\alpha b) = -B^{-1} < 0$. When $C' = 0$, the equation is

$$\frac{dv}{d\xi} = \frac{v}{2\alpha b}(v - 2u), \tag{9}$$

whose solution is

$$v = \frac{2u}{1 - \exp[(u\xi/\alpha b) + C]}. \tag{10}$$

Therefore, the kinetic energy density is

$$\varepsilon = \frac{1}{2}\rho v^2 = \frac{2\rho u^2}{\{1 - \exp[(u(r - ut)/b) + C]\}^2}. \tag{11}$$

When *r=0* and *t=0*, $\varepsilon = (2\rho u^2)/(1-e^c)^2$. Let the integral constant *c>0*, the larger the region is, the smaller the energy density $\varepsilon$ is at the same time t; the larger $\varepsilon$ becomes as time t increasing for the same region r. If $t_0 = (r/u) + c(4\eta + 3\zeta)/3\rho u^2$, $\varepsilon = \infty$. It is an unreachable value. Once the accumulation of energy ($\varepsilon$) with time at the same region excesses a faulting threshold value ($\varepsilon_0$) of rock in the locality, earthquake will occur. From this we may decide to have some relations among the geological structure, the faulting threshold value and the integral constant c.

The equation (9) corresponds a difference equation

$$v_{n+1} = \frac{1}{2\alpha b} v_n (v_n - 2u). \tag{12}$$

Using a substitution $v = u(1 - ux/2\alpha b)$, Eq.(12) becomes

$$x_{n+1} = 1 - \frac{u^4}{(2\alpha b)^2} x_n^2. \tag{13}$$

It is a well-known nonlinear chaos equation. For momentum the control parameter is

$$\lambda = (\frac{u^2}{2\alpha b})^2 = \frac{9u^4 \rho^2}{4\alpha^2 (4\eta + 3\xi)^2} > 0. \tag{14}$$

which determines the branch-chaos. If our consideration is in more detail, $\lambda$ will be more complex. Various similar results may be approximately derived from the energy conservation equations.

This fluid layer may be mantle, or a general asthenosphere. As an example, the matter density is $\rho \approx 4 g/cm^3$ and $\eta \approx 10^{22} p$, for the mantle. Let $\alpha \approx 1, t_0$ is enough long, $\lambda << 1$, so the usual crust is stable. But, if the mantle consists of solid, fluid, gas and plasma, $\eta$ will decrease very quickly. For a special asthenosphere, when $\lambda \geq 0.75$, the bifurcation will appear, perhaps it corresponds to the momentum migration between a couple of places. If $\eta$ decreases suddenly at a fault, $\lambda = 1.401155....$, it infers a chaos will appear, and corresponding earthquake occurs. In fact, earthquake will occur as the shorter the series of calm time, which is consistent with a branch-chaos process. This could be a basis of the renormalization schemes [10], which research some seismic activity prior to the main earthquake. In Webster's Third Dictionary "chaos"± possesses these meanings of chasm, gulf and abyss. These are namely results of earthquake. Some factors that cause earthquake, for example, volcanism, reservoir, underground nuclear explosion [15-17], etc., are merely boundary or initial conditions. But, because the seismic system obeys the nonlinear equation, they are extremely sensitive to the initial conditions.

The earthquake corresponds to chaos, which shows complexity on seismology, and is useful for the explanation of seismic essence, and casts a fatal shadow in an earthquake exact prediction. However, the possible periodicity in earthquakes may be determined according to the approximate results from a simplified model.

### III. Carlson-Langer model and magnitude-period formula

Combining the Carlson-Langer model [5-8], the movable asthenosphere corresponds to the surface, and the faulting rock layer corresponds to the block. It is consistent with geodynamics. From this the Gutenberg-Richter relation and the same results may be obtained, while in such case the masses of blocks are usually different, and the velocities may be changeable, especially at the fissure. Further, from some formulas of the Carlson-Langer model, the magnitude-period formula [18,19] may be derived. In the model, the magnitude *M* is be defined to be the natural logarithm of the earthquake moment *H*,

$$M = k \ln H. \tag{15}$$

Let the moment

$$H = C \int_{T_0}^{T+T_0} \dot{X} dt, \qquad (16)$$

between two times $T_0$ and $T_0 + T$. In the model, the frictional force is

$$F(\dot{X}) = \frac{F_0}{1+\dot{X}}, \qquad (17)$$

which is also a result in a simple mechanical model for earthquake dynamics [20]. In present paper the frictional force is

$$F = \eta \Delta V + (\frac{\eta}{3} + \zeta) grad div V \approx (\frac{4\eta}{3} + \zeta)v'' \\
= \frac{2u^3}{(\alpha b)^2} \exp(\frac{u\xi}{\alpha b} + c)[1 + \exp(\frac{u\xi}{\alpha b} + c)]/[1 - \exp(\frac{u\xi}{\alpha b} + c)]^3. \qquad (18)$$

If both forces F are equal,

$$\dot{X} = F_0 \frac{(\alpha b)^2}{2u^3} \frac{[1-\exp(u\xi/\alpha b + c)]^3}{\exp(u\xi/\alpha b + c)[1+\exp(u\xi/\alpha b + c)]} - 1. \qquad (19)$$

Only the time variable is concerned, so $\xi \approx -\alpha u t + c$, which is replaced to Eq.(16), and let $d = u^2/b$,

$$H_0 = C[F_0 \frac{(\alpha b)^2}{2u^3}(4\ln|1+e^{dt+c}| - e^{dt+c} - e^{-dt-c}) - t]_{T_0}^{T+T_0} \\
\approx C[F_0 \frac{(\alpha b)^2}{2u^3}(2 + 4dt + 4c) - t]_{T_0}^{T+T_0} = C(\frac{2\alpha^2 bF_0}{u} - 1)T. \qquad (20)$$

$$M = k \ln C + k \ln(\frac{2\alpha^2 bF_0}{u} - 1)T. \qquad (21)$$

Let $k \ln C = M_0$ and $(2\alpha^2 bF_0/u) - 1 = 1/T_0$, so

$$M = M_0 + k \ln \frac{T}{T_0} = M_0 + 2.3026 k \lg \frac{T}{T_0} = M_0 + \frac{1}{b}\lg(\frac{T}{T_0}). \qquad (22)$$

In this case, b is a tectonic parameter in the Gutenberg-Richter relation. The fractal characteristics of earthquakes show the fractal dimension D=2b [21]. If k=1, b=0.43429 agrees with the earthquake belt of the longer period of earthquakes, whose b=0.4-0.6. If k=0.5-0.333, b=0.86858-1.3029 [22]. The formula (22) is namely the magnitude-period formula of earthquakes [22]. This can be expressed in another form:

$$T = 10^{-b(M_0 - M)} T_0. \qquad (23)$$

The longer the period T of earthquakes is, the less the seismic number N is. The formula (23) and the Tsubokawa-Whitcomb-Scholz relation [23-25]

$$\log T = aM - b, \qquad (24)$$

between duration of crustal movement and magnitude of earthquake, have the same form and but different meanings and signs of parameters. The solutions of the one-dimensional partial differential equations for Carlson-Langer model have the form [8]:

$$U(x,t) \approx -(1-\varepsilon)\sin[\frac{x-vt}{\sqrt{v^2 - \xi^2}}], \qquad (25)$$

which implies a period.

Almost all formulas of the forecasts on earthquakes are based on the well-known Gutenberg-Richter relation, which can estimate the average rate of earthquakes. We assume that T and N are inversely proportional in a first-order approximation, so the magnitude-period formula (23) of earthquakes may be derived from the Gutenberg-Richter relation [18,19]. But, the formula is built on the nonlinear dynamical theory, so it may be researched widely.

When the magnetic field is neglected, and $\eta=0$, the hydrodynamic equations (1) combines the equation of continuity with the convection:

$$\rho_t + u\rho_x + w\rho_y = (N^2/g)w + k\Delta\rho. \qquad (26)$$

Then the Saltzman model and the well-known Lorenz model may be derived directly [26,27]. The Lorenz equations are:

$$dx/dt = -vx + ky, \qquad (27)$$
$$dy/dt = ax - by - xz, \qquad (28)$$
$$dz/dt = -cz + xy, \qquad (29)$$

where $x$ is the flow rate, $y$ and $z$ are asymmetric and symmetric parts of temperature difference on fluid, respectively. These are a simplified result of the Navier-Stokes equations. Usually, we suppose that all parameters are positive. If various parameters in Eqs.(27)(28)(29) take suitable values, one will obtain a beautiful Lorenz strange attractor in the nonlinear theory, which possesses two "wing". This corresponds perhaps to the earthquake migration to and fro, and two or a few time series on earthquakes in the same region.

According to the magnitude-period formula, so long as one supposes a possible period of earthquakes, for example, the period 34 year in California or the period 34 year in Romania, other periods can be calculated approximately. Bi and Yuan proposed a periodic scale 250 year for change of weather in China from various aspects [28]. We extended the periodic scale $T_0=250$ year to a period of earthquakes for $M_0=7$, and results are consistent with many known earthquakes in China [18]. From this the quantitative calculations can be given. When b=0.86868, various periods are T=33.83, 4.579, 0.6197 year for M=6, 5, 4, respectively. That T =33.83 year agrees with the periods of earthquakes in California, and in Romania. In the Romania region there are two series:

| Year | 1904 | 1935 | 1967 | 1999 |
|---|---|---|---|---|
| Magnitude | 7.5 | 6 | 7 | 7.8 |
| | 1916 | 1947 | 1980 | 2012 |
| | 6.5 | 7 | 6.2 | ? |

This may also be two different periods about 12 and 20 years: 1904+12+19+12+20+13+19+13. In the California region there is a series of period 32 years:

| Year | 1857 | 1890 | 1922 | 1954 | 1986 |
|---|---|---|---|---|---|
| Magnitude | 7.9 | 6.8 | 6.8 | 6.6 | 6.1 |

If b=1.70, the period will be T=4.988 year for M=6. This corresponds to the following earthquakes in California [28]:

| Year | 1918 | 1922 | 1926 | 1932 | 1937 | 1942 | 1947 | 1952 | 1956 |
|---|---|---|---|---|---|---|---|---|---|
| Magnitude | 6.8 | 6.8 | 6.1 | 6.4 | 6 | 6.5 | 6.4 | 7.7 | 6 |
| | 1962 | 1966 | 1971 | 1976 | 1981 | 1986 | 1991 | 1994 | 1999 |
| | 6.3 | 6.6 | 6.6 | 6.9 | 6.4 | 6.1 | 6.1 | 6.7 | 7.6 |

The period T that corresponds to the magnitude M will depend on b only, if the scale is determined. For example, according to the above table, in California next earthquake should take place in 2004 for T=4.988≅5. In fact, earthquakes (M=6.5) did occur on 23 December of 2003. According to the magnitude-period formula, in California next earthquake will take place in 2009, 2014 and 2019, etc. Further, we take the period T=33, a large earthquake of 2019 will be more possible [29,19]. The formula (23) has been discussed applying to the earthquakes in some regions in China. Either evaluation of T from b or evaluation of b from T agrees approximately with the facts. It is a simplified and calculable model.

Gerstenberger, et al., calculated the time-dependent seismic hazard by combining stochastic models derived from the Gutenberg-Richter relation and the modified Omori law, and proposed a clustering model, in which the rate at time t of aftershocks with magnitude $M \geq M_c$ is given by

$$\lambda(t) = 10^{a'+b(M_m-M)} /(t+c)^p . \qquad (26)$$

It possesses similar with our formula (23). Further, they discussed real-time forecasts of tomorrow's earthquakes in California [30].

If the conditions in the equations are considered in more detail and more complex, the magnitude-period formula will be revised. At some aspects, earthquake is analogous to the water leaking from the tap [31], in which energy accumulation for earthquake corresponds to weight accumulation for dripping, and the tension of crustal rock corresponds to the surface tension of the water. When the control parameter $\lambda$ reaches to some thresholds, one period will become periods two, four, eight, etc. It corresponds to two or a few time series, and to the earthquake migration to and fro in the same region. Even the randomness will appear owing to the sensitivity of the nonlinear system for the initial condition. But in this case earthquakes produce some new characters that possess the strange attractor and fractals. The Eq.(12) is extended to the complex functions, so it can become $Q(z) = z^2 + c$, whose fractal figure of Julia set is analogous to the distribution of earthquakes, specially, for c=i.

In the equations (1) if the external force $F_0$ includes some periodic interactions, such as sunspot, geomagnetic activity [32], influence of the celestial body [33], solar magnetic triggering mechanism [34], and crustal movement, etc., the earthquake as a sensitive nonlinear system will take place in the same periodicity probably, which is analogous to the forced vibration, and the space-time regions of earthquake will have some fluctuations.

Of course, this is a systematical and simplified model. If different conditions are concerned, the different equations will be derived. For instance, when the heat conduction and convection cannot be neglected, the thermal transport equation should be combined. For the electromagnetic phenomena in the earthquake, the more complex magnetohydrodynamics should be discussed. Therefore, the various links above may further be gradually developed and perfected. Moreover, the whole theory may be corrected and completed.

Resent, Ramos, et al., introduced a modification of the Olami-Feder-Christensen earthquake model in order to improve the Burridge-Knopoff mechanical model. Dynamical disorder is added to the thresholds following a narrow distribution. They found quasiperiodic behavior in the avalanche time series with a period proportional to the degree of dissipation of the system [35]. Periodicity is not as robust as criticality when the threshold force distribution widens, or when an increasing noise is introduced in the values of the dissipation.

Faces to huge natural suffering, if humanity could give up some conservative viewpoints, scientists explore bravely different methods: various modern scientific instruments, different scientific theories, and some paranormal ways [36], etc. A network of multilevel earthquake prediction will be able to be developed, and the accuracy of prediction will be increased.